%% file: MainPaper.tex
\documentclass[sigconf]{acmart}

\usepackage{booktabs} 
\usepackage{ccicons}  
\usepackage{todonotes}
\usepackage{subfig}
\graphicspath{{figures/}}
\usepackage[english]{babel}
\usepackage[autostyle=true,english=american]{csquotes}
\usepackage{gensymb}
\usepackage{balance}
\usepackage[most]{tcolorbox}
\usepackage{eurosym}
\usepackage{amsmath}
\usepackage{graphicx}
\usepackage{advdate}

\usepackage{acronym}
\newacro{AI}[AI]{Artificial Intelligence}
\newacro{UI}[UI]{user interface}
\newacro{GUI}[GUI]{graphical user interface}
\newacro{TLX}[TLX]{NASA-Task Load Index}
\newacro{RTLX}[Raw-TLX]{NASA Raw-Task Load Index}
\newacro{ER}[ER]{error rate}
\newacro{TCT}[TCT]{task completion time}
\newacro{HCI}[HCI]{Human-Computer Interaction}
\newacro{UX}[UX]{user experience}
\newacro{HFE}[HFE]{Human Factors and Ergonomics}
\newacro{cuDNN}[cuDNN]{CUDA Deep Neural Network library}
\newacro{RMSE}[RMSE]{root mean squared error}
\newacro{HMD}[HMD]{Head-Mounted Display}
\newacro{RF}[RF]{Random Forest}
\newacro{GP}[GP]{Gaussian process, long-plural = Gaussian processes}
\newacro{KNN}[\textit{k}NN]{\textit{k}-nearest neighbor}
\newacro{NN}[NN]{Neural Network}
\newacro{DNN}[DNN]{ Deep Neural Network}
\newacro{CNN}[CNN]{Convolutional Neural Network}
\newacro{FCL}[FCL]{fully connected layer}
\newacro{BoD}[BoD]{Back-of-Device}
\newacro{FOV}[FoV]{field of view}
\newacro{RW}[RW]{Real World}
\newacro{IFRC}[IFRC]{index finger ray cast}
\newacro{FRC}[FRC]{forearm ray cast}
\newacro{EFRC}[EFRC]{eye-finger ray cast}
\newacro{HRC}[HRC]{Human-Robot Collaboration}
\newacro{HRI}[HRI]{Human-Robot Interaction}
\newacro{6DOF}[6DOF]{six-degree-of-freedom}
\newacro{LOOCV}[LOOCV]{leave-one-out cross-validation}
\newacro{CV}[CV]{cross-validation}
\newacro{RM}[RM]{repeated measure}
\newacro{ANOVA}[ANOVA]{analysis of variance}
\newacro{RMANOVA}[RM-ANOVA]{repeated measures analysis of variance}
\newacro{AGATe}[AGATe]{AGreement Analysis Toolkit}
\newacro{GHoST}[GHoST]{Gesture Heatmap Toolkit Gesture Heatmaps Toolkit}
\newacro{GREAT}[GREAT]{Gesture Relative Accuracy Toolkit}
\newacro{GRT}[GRT]{Gesture Recognition Toolkit}
\newacro{DTW}[DTW]{Dynamic Time Warping}
\newacro{LHRD}[LHRD]{large high resolution display}
\newacro{GEQ}[GEQ]{Game Experience Questionnaire}
\newacro{SPGQ}[SPGQ]{Social Presence Gaming Questionnaire}
\newacro{JND}[JND]{just-noticeable difference}
\newacro{SUS}[SUS]{system usability scale}
\newacro{CSCW}[CSCW]{computer-supported cooperative work}
\newacro{CAD}[CAD]{computer-aided design}
\newacro{MR}[MR]{Mixed Reality}
\newacro{CVE}[CVE]{Collaborative Virtual Environment}
\newacro{AR}[AR]{Augmented Reality}
\newacro{AV}[AV]{Augmented Virtuality}
\newacro{VR}[VR]{Virtual Reality}
\newacro{PRISMA}[PRISMA]{Preferred Reporting Items for Systematic Reviews}
\newacro{PRISMA-Scope}[PRISMA-ScR]{Meta-Analyses Extension for Scoping Reviews}
\newacro{TF-IDF}[TF-IDF]{Term Frequency-Inverse Document Frequency}
\newacro{TF}[TF]{Term Frequency}
\newacro{AVs}[AVs]{Automated Vehicles}
\newacro{eHMIs}[eHMIs]{external Human-machine interfaces}
\newacro{SAR}[SAR]{Spatial Augmented Reality}
\newacro{IFR}[IFR]{International Federation of Robotics}
\newacro{ADLs}[ADLs]{Activities of Daily Living}
\newacro{LED}[LED]{Light-Emitting Diode}
\newacro{DoF}[DoF]{Degrees-of-Freedom} \newacroplural{DoF}[DoFs]{Degrees-of-Freedom}
\newacro{HHC}[HHC]{Human-Human Collaboration}
\newacro{IDF}[IDF]{Inverse Document Frequency}
\newacro{ML}[ML]{Machine Learning}
\newacro{HMM}[HMM]{hidden Markov model}
\newacro{IK}[IK]{inverse kinematics}

\AtBeginDocument{%
  \providecommand\BibTeX{{%
    \normalfont B\kern-0.5em{\scshape i\kern-0.25em b}\kern-0.8em\TeX}}}

\copyrightyear{2024}
\acmYear{2024}
\setcopyright{rightsretained}
\acmConference[HRI '24 Companion]{Companion of the 2024 ACM/IEEE International Conference on Human-Robot Interaction}{March 11--14, 2024}{Boulder, CO, USA}
\acmBooktitle{Companion of the 2024 ACM/IEEE International Conference on Human-Robot Interaction (HRI '24 Companion), March 11--14, 2024, Boulder, CO, USA}
\acmDOI{10.1145/3610978.3640635}
\acmISBN{979-8-4007-0323-2/24/03}

\makeatletter
\gdef\@copyrightpermission{
  \begin{minipage}{0.3\columnwidth}
   \href{https://creativecommons.org/licenses/by/4.0/}{\includegraphics[width=0.90\textwidth]{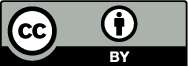}}
  \end{minipage}\hfill
  \begin{minipage}{0.7\columnwidth}
   \href{https://creativecommons.org/licenses/by/4.0/}{This work is licensed under a Creative Commons Attribution International 4.0 License.}
  \end{minipage}
  \vspace{5pt}
}
\makeatother

\begin{document}

\title{Hands-On Robotics: Enabling Communication Through Direct Gesture Control}

\author{Max Pascher}
\orcid{0000-0002-6847-0696}
\email{max.pascher@udo.edu}
\affiliation{
    \institution{TU Dortmund University}
    \city{Dortmund}
    \country{Germany}
}

\author{Alia Saad}
\orcid{0000-0002-9910-295X}               
\email{alia.saad@uni-due.de}
\affiliation{
    \institution{University of Duisburg-Essen}
    \city{Essen}
    \country{Germany}
}

\author{Jonathan Liebers}
\orcid{0000-0002-6923-9066}
\email{jonathan.liebers@uni-due.de}
\affiliation{
    \institution{University of Duisburg-Essen}
    \city{Essen}
    \country{Germany}
}

\author{Roman Heger}
\orcid{0000-0001-6705-2145}               
\email{roman.heger@uni-due.de}
\affiliation{
    \institution{University of Duisburg-Essen}
    \city{Essen}
    \country{Germany}
}

\author{Jens Gerken}
\orcid{0000-0002-0634-3931}
\email{jens.gerken@udo.edu}
\affiliation{
    \institution{TU Dortmund University}
    \city{Dortmund}
    \country{Germany}
}

\author{Stefan Schneegass}
\orcid{0000-0002-0132-4934}
\email{stefan.schneegass@uni-due.de}
\affiliation{
    \institution{University of Duisburg-Essen}
    \city{Essen}
    \country{Germany}
}

\author{Uwe Gruenefeld}
\orcid{0000-0002-5671-1640}
\email{uwe.gruenefeld@uni-due.de}
\affiliation{
    \institution{University of Duisburg-Essen}
    \city{Essen}
    \country{Germany}
}

\renewcommand{\shortauthors}{Max Pascher et al.}
\newcommand{\alia}[1]{\textsf{\textcolor{magenta}{\textbf{Alia:} \textit{#1}}}}
\newcommand{\maxi}[1]{\textsf{\textcolor{blue}{\textbf{Max:} \textit{#1}}}}
\begin{abstract}
\input{content/00-abstract}
\end{abstract}



\begin{teaserfigure}
  \includegraphics[width=\textwidth]{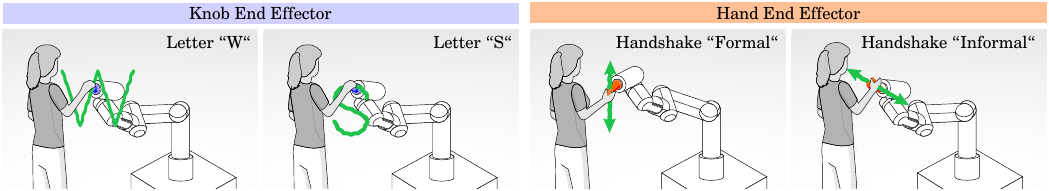}
  \caption{We investigated four different user performed gestures by directly manipulating a robot; two letter-based gestures using a knob as an end effector; and two handshake gestures with a hand as an end effector.}
  \Description[Teaser figure]{The figure consists of four subfigures, each showing an illustration of a human grasping the end effector of a robotic arm performing a gesture. In both subfigures on the left, a knob is mounted to the end effector for a letter-based gesture ('W' and 'S'), and the subfigures on the right side show a hand mounted to the end effector for 'formal' and 'informal' handshake gestures.}
  \label{fig:teaser}
\end{teaserfigure}

\maketitle

\input{content/01-introduction}
\input{content/02-related-work}

\input{content/03-approach}
\input{content/04-study}
\input{content/05-ml}
\input{content/06-discussion}
\input{content/07-summary-and-conclusion}

\begin{acks}
This work is funded by the \emph{Deutsche Forschungsgemeinschaft (DFG, German Research Foundation)} -- Project ID \href{https://gepris.dfg.de/gepris/projekt/425869382}{425869382} \& \href{https://gepris.dfg.de/gepris/projekt/426052422}{426052422} and the \textit{German Federal Ministry of Education and Research} (BMBF, FKZ: \href{https://foerderportal.bund.de/foekat/jsp/SucheAction.do?actionMode=view&fkz=16SV8565}{16SV8565}).
\end{acks}

\bibliographystyle{ACM-Reference-Format}
\bibliography{MainPaper}

\end{document}

%% file: content/00-abstract.tex

Effective \ac{HRI} is fundamental to seamlessly integrating robotic systems into our daily lives. However, current communication modes require additional technological interfaces, which can be cumbersome and indirect. This paper presents a novel approach, using direct motion-based communication by moving a robot's end effector. Our strategy enables users to communicate with a robot by using four distinct gestures -- two handshakes ('formal' and 'informal') and two letters ('W' and 'S'). As a proof-of-concept, we conducted a user study with 16 participants, capturing subjective experience ratings and objective data for training machine learning classifiers. Our findings show that the four different gestures performed by moving the robot's end effector can be distinguished with close to 100\% accuracy. Our research offers implications for the design of future \acs{HRI} interfaces, suggesting that motion-based interaction can empower human operators to communicate directly with robots, removing the necessity for additional hardware.


%% file: content/01-introduction.tex
\section{Introduction}
\label{sec:introduction}

In recent years, rapid advancements in robotic technologies have led to their growing integration in our daily lives, acting as versatile assistants in workplaces and homes~\cite{Pascher.2021, Ajoudani.2017.collaboration, Mahdi.2022}. These robotic companions enhance human capabilities and efficiency and, as such, substantially change how we interact with the world~\cite{Galin.2020}. As robotic solutions evolve and diversify, their capacity of autonomous actions increases -- with seamless close-contact interactions between humans and robots becoming a reality~\cite{Cherubini2016}.

However, for successful \ac{HRC}, effective communication channels must be established for accurate transmission of intent and coordination of respective actions~\cite{Pascher.2023robotMotionIntent}. Traditional modes of \ac{HRI}, such as voice commands~\cite{Badr2020} or touch interfaces~\cite{Andreasson2017} can effectively convey instructions to robots. Yet, as robots take on more complex tasks and work in proximity to humans -- sometimes working \emph{hand in hand} -- the need for a more natural and intuitive communication approach becomes apparent. 
\emph{Motion-based communication}, where humans actively manipulate the robot's end effector, can be a viable solution to this challenge.
By mimicking how we naturally interact with one another, \emph{motion-based communication} bridges the gap between humans and robots, promoting a more intuitive and seamless exchange of information and -- consequently -- improved collaboration.





Therefore, tracking and measuring movements is a crucial aspect to consider. Two types of tracking approaches -- \textit{relative} and \textit{absolute}-- exist. Inertial sensors (e.g., accelerometers~\cite{bao2004activity}) provide relative tracking information, while optical tracking methods (e.g., cameras~\cite{abawi2004accuracy}) offer more exact positional data.
Previous studies, like those using a \emph{Nintendo Wii} controller~\cite{Schlomer.2008wiicontroller}, have successfully demonstrated gesture recognition using inertial sensors with limited training samples. In contrast, our approach focuses on mechanical tracking -- positioned between \textit{relative} and \textit{absolute} tracking -- using a seven \acp{DoF} robotic arm for precise movement data. Our research uses the robot's joint motion data to enable accurate and efficient \emph{motion-based communication}.

Gesture recognition has long been recognized as a challenging yet crucial aspect of enabling natural and intuitive communication between humans and machines. Recognizing gestures using pattern recognizers from the \$-family has been one of the earlier approaches in the field~\cite{Wobbrock.2007dollar1, Anthony.2010dollarN, Anthony.2012dollarNprotractor, Vatavu.2012dollarP, Vatavu.2017dollarPplus, Vatavu.2018dollarQ}, leading to numerous follow-ons by other researchers~\cite{Magrofuoco2021,LI2019,Echtler2012,Juli2013}. 
These recognizers utilize a predefined set of gesture templates to match and identify user motions. Template-based solutions showed robust performance both in 2D and 3D spaces~\cite{Caramiaux2014}. 

\ac{ML} techniques are potent tools for gesture recognition~\cite{Nogales2021}, allowing systems to learn and adapt to a wide range of gestures. Although it may seem intuitive that existing \ac{ML} approaches will work for recognizing gestures directly performed with a robot's end effector, so far, this remains unproven. 
When designing gestures, \textit{affordance} plays a significant role in determining the gesture vocabulary. An artificial hand on a robot may prompt gestures like a handshake, while the affordance of a knob encourages other gestures.

This work uses the robot's sensor data to study recognizing and distinguishing gestures performed by direct interaction with a robotic arm. We performed a laboratory study (N=16) to assess the feasibility of four distinct gestures and determine their respective recognition accuracy, resulting in a f1-score up to 0.99. Here, we demonstrate the feasibility of direct interactions' usage for robust gesture recognition in \ac{HRI}. 

%% file: content/02-related-work.tex
\section{Related Work}
Our research integrates insights from collaborative robots in close-contact interactions and embodied user input for gesture recognition and classification. 

In \ac{HRC}, collaborative robots -- known as cobots -- are increasingly common in various settings, including domestic care~\cite{Bemelmans.2012, Pascher.2021}. They are categorized based on environment sharing~\cite{shi2012levels, pini2015systematic, michalos2015design} and types of cooperation~\cite{helms2002rob, bauer2016lightweight}. Previous research focused on cobots adapting to human movements and behavior, while maintaining appropriate distance~\cite{mumm2011}, avoiding collisions~\cite{hoang2019}, and customizing assistance based on skills~\cite{Blanchet.2020} and comfort~\cite{Chen.2018.P}. 
Supporting this,\citeauthor{drolshagen2021} indicated no adverse effects on collaboration when safety aspects are met~\cite{drolshagen2021}, while \citeauthor{maurtua2017human} questioned study participants that anticipate increased interaction with cobots in the future~\cite{maurtua2017human}.
Efficiency in \ac{HRC} can be enhanced with suitable techniques and sensors~\cite{Bonci.2021}. 
The embodiment of cobots positively affects perception and trust~\cite{Ye.2020}, while touch-based interactions improve non-verbal communication and reduce human stress responses~\cite{Willemse.2019}. Interactive perception combines physical and traditional methods but may be limited by occlusion~\cite{kragic2018interactive}.


\textit{Gesture recognition} plays a pivotal role in \ac{HRI}. 
Utilizing the acceleration sensor of the \emph{Nintendo Wiimote}, \citeauthor{Schlomer.2008wiicontroller} implemented gesture recognition by employing a \ac{HMM} for training and recognizing user-selected gestures~\cite{Schlomer.2008wiicontroller}. Despite the small training set, their evaluation demonstrated an accuracy between 0.85 to 0.95. \citeauthor{Wu2009} proposed a similar approach using acceleration-based movement data, achieving an accuracy of almost 0.99 for four gestures using the \emph{FDSVM} method. \citeauthor{Cabrera2016} used the \emph{Microsoft Kinect} sensor to use skeleton data for one-shot gesture recognition~\cite{Cabrera2016}, comparing three classifiers with accuracy between 0.81 and 0.86.  Using a WiFi-based approach, gesture recognition for multi-user applications was demonstrated by \citeauthor{Venkatnarayan2021}~\cite{Venkatnarayan2021}. Their system identifies concurrent gestures, quantifies the total gesture number, and creates virtual samples for different combinations using training data from a single user. Achieving an accuracy of over 0.90, it can recognize up to eight gestures performed simultaneously.


By leveraging knowledge and techniques from collaborative robotics and gesture recognition, advanced systems capable of accurate gesture recognition --  while adapting and responding to human movements and behavior -- become possible. These capabilities result in smoother and more efficient \ac{HRC}, fostering overall safer and more productive interactions.

%% file: content/03-approach.tex
\section{Gesture Input Interactions}

We investigate the use of a robotic arm for effective embodied gesture recognition, specifically focusing on movements that directly manipulate the robot's end effector through physical interaction.  
In domestic care, assistive robotic arms often use multi-finger end effectors comparable to a human hand. For this kind of gripper, we use two types of handshakes as natural gestures. The robot's flanch also supports adding a grasping object (e.g., a protrusion).






Several gestures have been introduced in prior works~\cite{huang2021robot}. Here, we selected two \textbf{letter-based gestures} and two \textbf{handshake gestures}, simulating a natural human-like interaction with the robotic arm (see~\autoref{fig:teaser}).
To accommodate the differing movements and grips between the gestures (letters vs. hand), we attached a spherical knob for the letters, enabling a firmer grip and smoother motion in 3D space. A hand model resembling a human hand was used for a more lifelike interactions.

\begin{description}
    \item[Letter] Using the knob end effector, we investigated two types of motions: \emph{circular}, represented by the curved letter \enquote{S} fostering a smooth transition in performing the gesture, and \emph{linear}, represented by the sharp-edged letter \enquote{W} consisting of several stops for direction changing during the motion.
    \item[Handshake] This gesture imitates the traditional human-to-human handshake and, as such, was performed with the hand end effector. Potential applications include an introductory interaction to initiate communication with a cobot to start a procedure. We examined both the \emph{formal} handshake, involving grasping the hand followed by an up-and-down movement, and the \emph{informal} one, also known as the \emph{G-lock handshake}.
\end{description}



    
    
    
    

 


%% file: content/04-study.tex
\section{Study: Gesture Recognition Using a Robotic Arm}
In this study, we introduce a novel approach in human-to-robot communication through gesture-based mechanical manipulation of a robotic arm. Our method centers on extracting movement values from the robotic arm's joints and does not depend on additional tracking requirements. We investigate how mechanical manipulation of a robotic arm can achieve accurate gesture recognition.


\subsection{Study Design}
We conducted a within-subject controlled laboratory study to assess the accuracy of gesture recognition.
The independent variable were \emph{gestures}, with two pre-selected letter gestures (denoted \emph{LS} and \emph{LW}) and two handshake gestures (denoted \emph{HS} and \emph{GL}). The \emph{gesture recognition accuracy} serves as the dependent variable. 

\subsection{Participants and Procedure}
We recruited 16 participants (6 females, 10 males), aged between 22 and 35 years(M$=$27.75, SD$=$3.96) via mailing lists and social media.
All participants were right-handed and reported no motor functions limitations or injuries. The study received approval from our institution's ethics committee, and each participant received a 10 Euro remuneration upon task completion.

Study sessions for each participant began with the experimenter explaining the study's purpose and demonstrating the interaction process. Subsequently, participants completed demographic and consent forms.
At the start of each run, the robot arm end effector was automatically positioned to a predefined starting point to maintain consistency across all participants.
The study conductor briefed participants on the interaction pattern, allowing them to start and stop freely. The conductor then started the recording of the robot's movement. The corresponding movement data were recorded while participants were moving the robot arm end effector in the desired gesture. Each gesture was recorded five times consecutively, totaling 20 recordings per participant. Gesture order was counterbalanced using a Latin Square design. 
Following the completion of all recordings, a post-study questionnaire was administered. On average, participants finished the entire task within 40 minutes.


\subsection{Apparatus}
For our study, we used a Franka Emika robot placed on a fixed table at a height of 61\,cm~\cite{franka-emika-robot}.
Both types of end effectors were 3D printed by the research team; the models were acquired from an open-source library~\cite{thingiverse-open-source-library}.
Manipulating the robotic arm required setting the robot status to the \emph{free guiding} mode for a safe interaction. We affixed a clamp to engage the \emph{free guiding} mode buttons, allowing users to freely execute gestures. This action also deactivated any ongoing autonomous movements by the robot.
While performing the gestures, participants did not receive any additional feedback.



\subsection{Results}
\begin{figure*}%
    \centering
    \subfloat{{\includegraphics[width=0.2\linewidth]{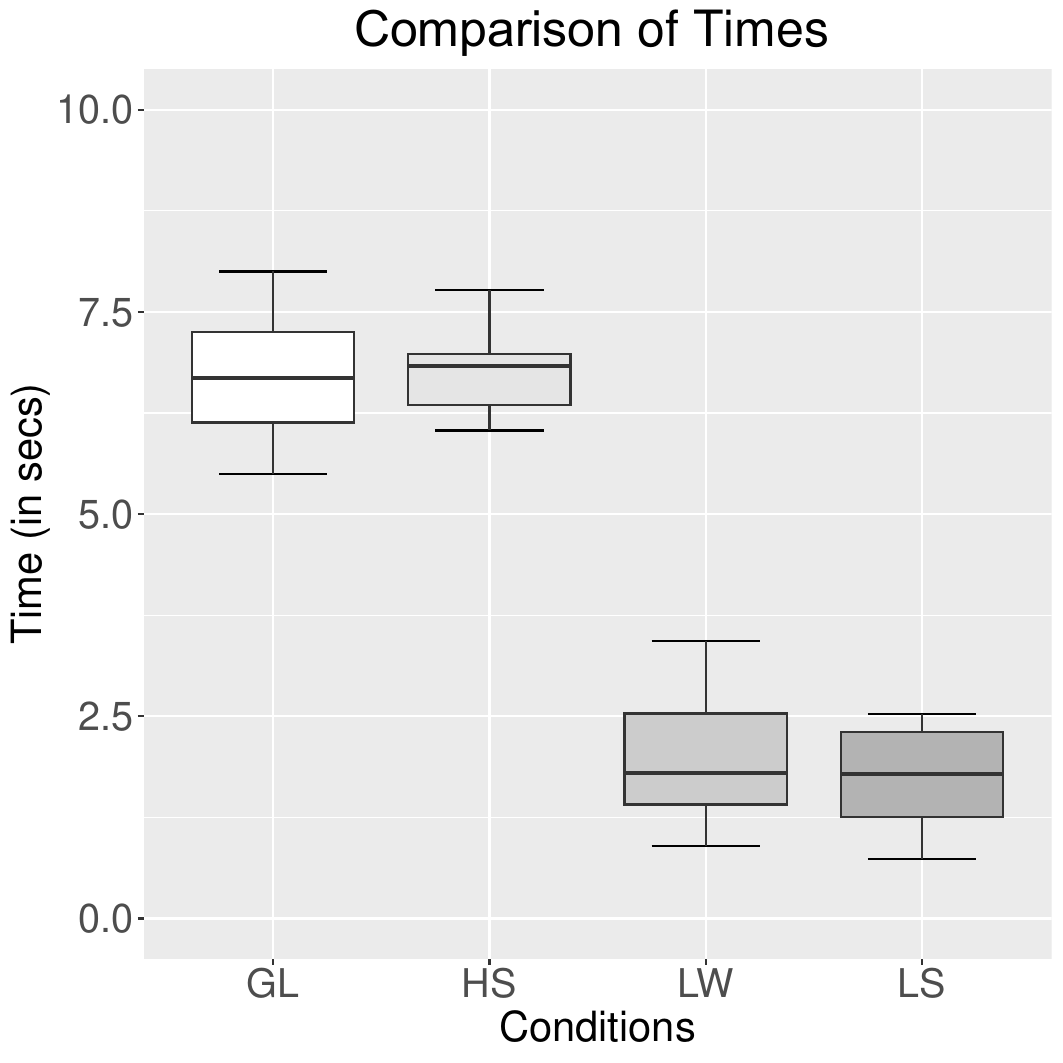} }}%
    \hfill
    \subfloat{{\includegraphics[width=0.2\linewidth]{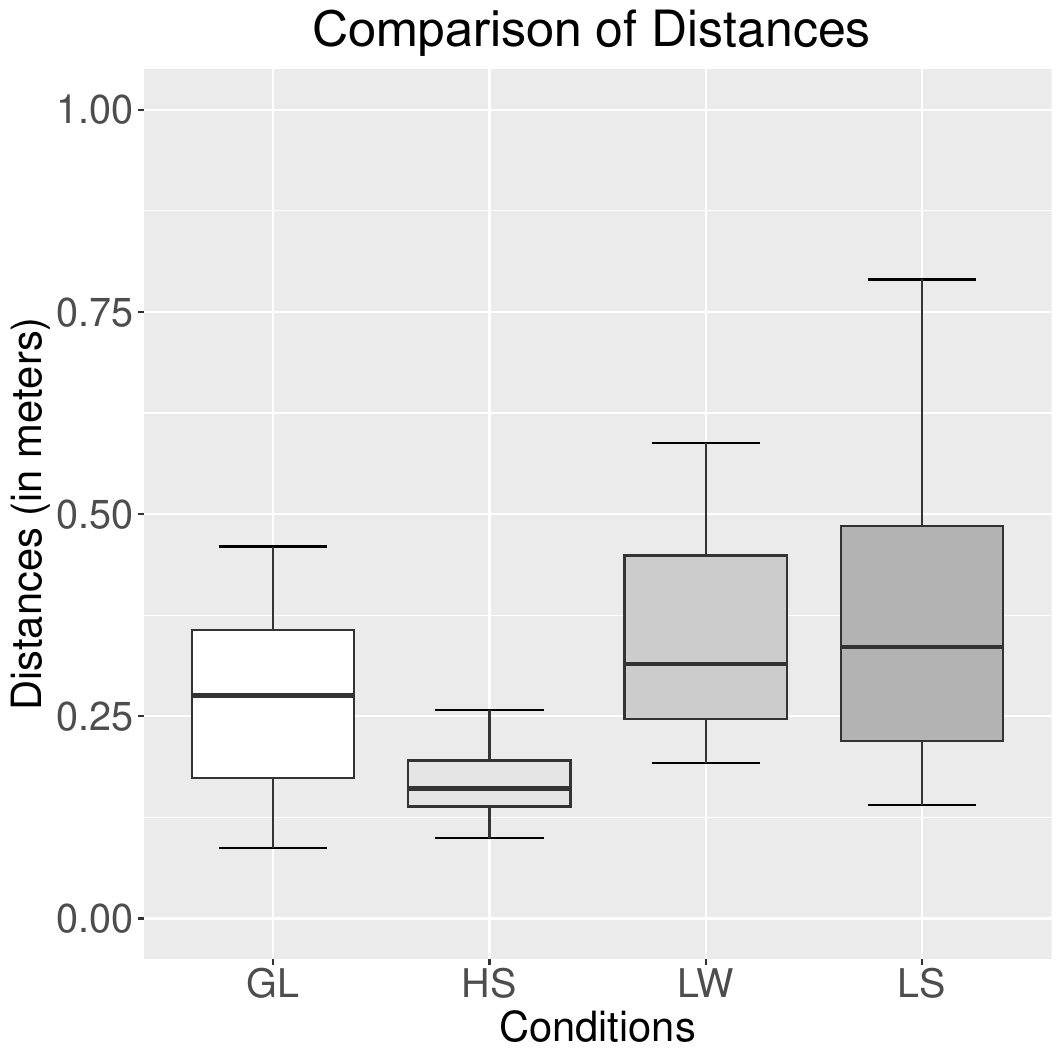} }}%
    \hfill
    \subfloat{{\includegraphics[width=0.235\linewidth]{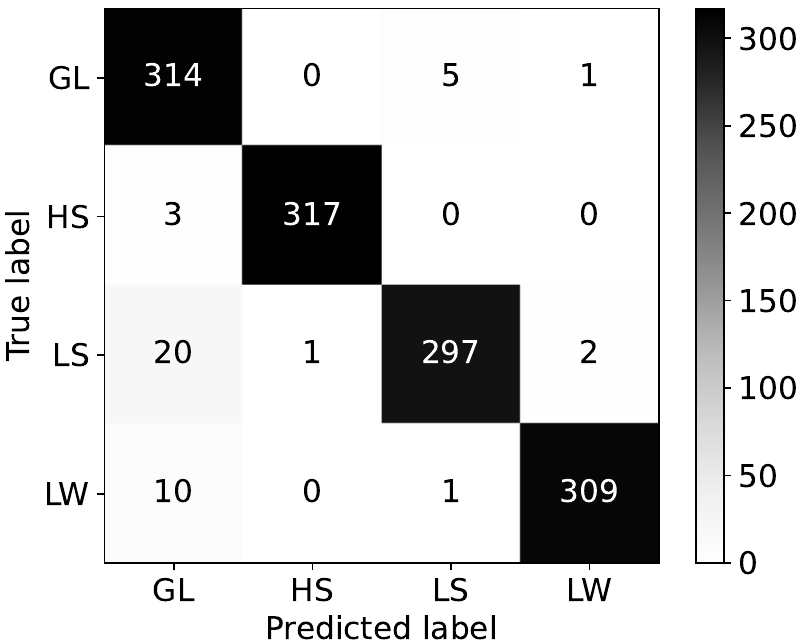} }}%
    \hfill
    \subfloat{{\includegraphics[width=0.23\linewidth]{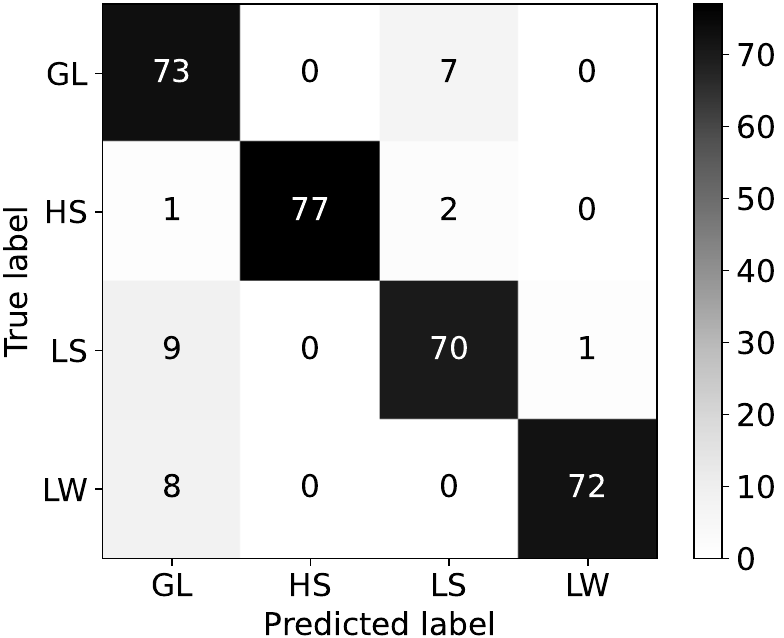} }}%
    \caption{Measures of gestures, for \textbf{(first)} temporal, \textbf{(second)} spatial dimensions, and confusion matrices, for \textbf{(third)} inverse classification (cross-validation), the training sample~$=$~0.2, and \textbf{(fourth)} user-independent (cross-subject), with training sample~$=$~0.5.}
    \Description[Results]{The figure consists of four subfigures. The two subfigures on the left represent boxplots reporting the 'comparison of time' and 'comparison of distances' between all four gestures. The two subfigures on the right show confusion matrices for 'inverse classification' (cross-validation) and 'user-independent' (cross-subject) classification.}
    \label{fig:confusion_matrices}%
\end{figure*}
In our analysis, we collected \textit{objective} and \textit{subjective} measures. We applied non-parametric Friedman tests to detect significant main effects between gestures. Post-hoc, we conducted Wilcoxon signed-rank tests (Bonferroni corrected) for pairwise comparisons. The effect sizes of the Wilcoxon tests are reported as r (r: $>$0.1 small, $>$0.3 medium, and $>$0.5 large effect).

\subsubsection{Objective Measures}
For objective measures, we report the median (interquartile range) of the \textit{temporal} and \textit{spatial} dimensions of each gesture in ascending order (see~\autoref{fig:confusion_matrices} left).

\textbf{Temporal:}
The duration per gesture are: \emph{LS}$=$1.78s (IQR$=$1.05s), \emph{LW}$=$1.80s (IQR$=$1.12s), \emph{GL}$=$6.68s (IQR$=$1.12s), and \emph{HS}$=$6.83s (IQR$=$ 0.63s).
A Friedman test revealed significant differences between the conditions ($\chi^2$(3)$=$39.00, p$\leq$0.001, N$=$16).
Post-hoc tests showed significant differences between \emph{GL} and \emph{LW} (W$=$136, Z$=$3.52, p$\leq$0.001, r$=$0.62), \emph{GL} and \emph{LS} (W$=$136, Z$=$3.52, p$\leq$0.001, r$=$0.62), \emph{HS} and \emph{LW} (W$=$136, Z$=$3.52, p$\leq$0.001, r$=$0.62), and \emph{HS} and \emph{LS} (W$=$136, Z$=$3.52, p$\leq$, r$=$0.62), but no significant differences between \emph{GL} and \emph{HS} (W$=$62, Z$=$-0.31, p$=$1) and \emph{LW} and \emph{LS} (W$=$62, Z$=$-0.31, p$=$1).

\textbf{Spatial:}
The maximum distance per gesture are: \emph{HS}$=$0.16m (IQR$=$0.06m), \emph{GL}$=$0.28 (IQR$=$0.18m), \emph{LW}$=$0.31 (IQR$=$0.20m), and \emph{LS}$=$0.34m (IQR$=$0.27m).
A Friedman test revealed significant differences between the conditions ($\chi^2$(3)$=$21.38, p$\leq$0.001, N$=$16).
Post-hoc tests showed significant differences between \emph{HS} and \emph{LW} (W$=$0, Z$=$-3.52, p$\leq$0.001, r$=$0.62) and \emph{HS} and \emph{LS} (W$=$0, Z$=$-3.52, p$\leq$0.001, r$=$0.62), but no significant differences between \emph{GL} and \emph{HS} (W$=$116, Z$=$2.48, p$=$0.066), \emph{GL} and \emph{LW} (W$=$37, Z$=$-1.60, p$=$0.700), \emph{GL} and \emph{LS} (W$=$36, Z$=$-1.65, p$=$0.627), and \emph{LW} and \emph{LS} (W$=$48, Z$=$-1.03, p$=$1).


\subsubsection{Subjective Measures}
We used the Likert-items for \textit{Mental Demand} and \textit{Physical Demand} of the NASA Raw-TLX questionnaire~\cite{hart1988development} to assess subjective measures regarding task load. The median task load scores (with interquartile ranges) for each gesture, in ascending order, are as follows:
For \textbf{Mental Demand} we found \emph{LW}$=$10.00 (IQR$=$5.00), \emph{LS}$=$10.00 (IQR$=$11.25), \emph{HS}$=$5.00 (IQR$=$5.00), and \emph{GL}$=$7.50 (IQR$=$5.00). A Friedman test revealed no significant differences between the conditions ($\chi^2$(3)$=$5.23, p$=$0.156, N$=$16).
For \textbf{Physical Demand} we obtained \emph{LW}$=$10.00 (IQR$=$1.25), \emph{LS}$=$10.00 (IQR$=$7.50), \emph{HS}$=$10.00 (IQR$=$7.50), and \emph{GL}$=$12.50 (IQR$=$10.00). A Friedman test revealed no significant differences between the conditions ($\chi^2$(3)$=$1.04, p$=$0.792, N$=$16)



\textbf{Suggested Gestures:}
Participants were invited to suggest new gestures, and most (n$=$10) recommended knob-based gestures. These suggestions ranged from additional letters such as \enquote{I}, \enquote{L}, or \enquote{U}, to straightforward non-letter gestures like a straight line \enquote{---}, a checkmark \enquote{$\checkmark$}, and more intricate gestures like a question mark \enquote{?} or a square \enquote{$\square$}.
Certain suggestions leaned towards human-like interactions, including fist bumps, holding hands, or shaking objects. This aligns with participants' impressions of the interaction, describing the handshake as \enquote{natural} (P5, P6).




%% file: content/05-ml.tex
\section{Gesture Recognition}
Applying \ac{ML}-based gesture recognition, we conducted a cross-validated classification using an 80-20 and 20-80 training-testing data set and evaluated a user-independent classification.

\paragraph{Data Analysis}
For the training process, we computed four low-level descriptive statistical values (minimum, maximum, mean, and standard deviation) for all positions, as well as velocity and effort values per joint for each of the seven joints. This combination resulted in a set of 84 features per gesture, per participant.
As a classifier, we used the \ac{RF} implementation of \emph{scikit-learn} with default parameters (e.g., 100 estimators)~\cite{sklearn-rf-default-params}.

\paragraph{Cross-Validation Classification}
To analyze the effect of differences \emph{between gestures}, we calculated the mean accuracy for each individual gesture using k-fold cross-validation (k$=$5). We performed the classical 80-20 split of training and test set, and achieved an average f1-score of 0.99. To further test the robustness of distinguishing gestures across participants, we performed an inverse classification, where the train and test sets were 20\% and 80\%, respectively. An average f1-score of 0.96 was achieved. The confusion matrix of the 20-80 classifier is illustrated in~\autoref{fig:confusion_matrices} (right). 

\paragraph{User-Independent Classification}
To test the generalizability of our approach, i.e., its performance against new and unknown users, 
the data set was equally divided into disjoint training and test sets (i.e., with no overlap of participants). Classifying two folds achieved a mean f1-score accuracy of 0.91 (0.85 and 0.97 respectively).








%% file: content/06-discussion.tex
\section{Discussion} 
We investigated gesture recognition by using motion data of a robotic arm manipulated by the user, by incorporating \textit{relative} and \textit{absolute} tracking data of the robot's end effector. A critical consideration is the robot's \ac{IK}, which may deviate from natural human motion due to joint angle constraints. 

\paragraph{Validity of Mechanical Interaction with a Robot}
Cross-validation produced promising results, with varying train and test set sizes achieving f1-scores between 0.96 and 0.99. These outcomes highlight the potential for solid performance even with limited training data. Furthermore, user-independent classification demonstrated robustness in recognizing gestures, irrespective of users, achieving an accuracy rate of 0.91. These findings underscore the practicality of our solution across diverse users and scenarios, particularly in systems requiring gesture recognition from different individuals.



\paragraph{Subjective Feedback}

Participants perceived the gestures as neither \emph{mentally} nor \emph{physically} demanding, indicating a convenient and natural interaction. However, this perception might be influenced by the limited number of repetitions (N$=$5/gesture). Possible applications of this method include interrupting an undesired ongoing task or altering direction through a single, clear, and unambiguous command.

\paragraph{Limitations \& Future Work}
We selected and assessed various gestures with different handles for the robot's end effector, though not exhaustively exploring every combination of gesture and handle. Furthermore, our repertoire of gestures was limited to just four distinct ones.

Future work will involve expanding the number of repetitions per gesture and incorporating additional gestures. We anticipate conducting further studies based on this work, potentially involving a more diverse range of user groups. This may also encompass studies exploring the impact of altering robot tasks through mechanical input.


%% file: content/07-summary-and-conclusion.tex
\section{Conclusion}

We highlight mechanical-based gestures in \ac{HRI} by manipulating two robot end effectors (knob/hand) with four gestures (two letters / two handshakes). Classification achieved 0.91 for user-independent and 0.99 using 80\% of the collected data for training. This robust gesture recognition establishes mechanical interaction with the robotic arm as a feasible, immediate, and intuitive user input.
